\documentclass{article}
\usepackage[dvipsnames, svgnames, x11names]{xcolor}
\usepackage[utf8]{inputenc} 
\usepackage[T1]{fontenc}    
\usepackage{hyperref}       
\usepackage{url}            
\usepackage{booktabs}       
\usepackage{amsfonts}       
\usepackage{nicefrac}       
\usepackage{microtype}      
\usepackage{amsmath}
\usepackage{amssymb}
\usepackage{stmaryrd}
\usepackage{amsthm}
\usepackage{cleveref}
\usepackage{graphicx} 
\usepackage{caption}
\usepackage{mwe}
\usepackage{graphics,graphicx,epsfig,psfrag,subfigure,threeparttable,url}
\usepackage{float}
\usepackage{algorithm}
\usepackage{algorithmic}

\usepackage{color}
\usepackage{authblk}
\usepackage{amsmath,graphicx}

\title{SAM: A Self-adaptive Attention Module for Context-Aware Recommendation System}
\date{}
\author[1]{Jiabin Liu\thanks{Jiabin Liu and Zheng Wei contributed equally to this work.}}
\author[1]{Zheng Wei$^*$}
\author[1]{Zhengpin Li}
\author[1]{Xiaojun Mao\thanks{Xiaojun Mao and Jian Wang are the co-corresponding authors. }}
\author[1]{Jian Wang$^\dagger$}
\author[1]{Zhongyu Wei}
\author[2]{Qi Zhang}
\affil[1]{School of Data Science, Fudan University, Shanghai, China}
\affil[2]{School of Computer Science, Fudan University, Shanghai, China}
\begin{document}
\maketitle
\begin{abstract}
Recently, textual information has been proved to play a positive role in recommendation systems. However, most of the existing methods only focus on representation learning of textual information in ratings, while potential selection bias induced by the textual information is ignored. In this work, we propose a novel and general self-adaptive module, the Self-adaptive Attention Module (SAM), which adjusts the selection bias by capturing contextual information based on its representation. This module can be embedded into recommendation systems that contain learning components of contextual information. Experimental results on three real-world datasets demonstrate the effectiveness of our proposal, and the state-of-the-art models with SAM significantly outperform the original ones.
\end{abstract}

\section{Introduction}
With the explosive growth of data volume, recommendation systems~(RS) have been a prevailing and powerful tool for helping people to alleviate information overload. As the most commonly used collaborative filtering technique for RS, matrix factorization~(MF)~\cite{koren2009matrix, koren2008factorization} models the interactions of users and items behind historical ratings by learning a shared latent space for their representations. To date, many state-of-the-art methods have been developed based on MF, such as Probabilistic Matrix Factorization, Deep Matrix Factorization and Neural Collaborative Filtering. However, data sparsity and intrinsic bias of observations can affect the validity of latent presentations, thereby deteriorating the prediction accuracy.

To build more effective MF models, one mainstream solution is to utilize auxiliary textual information, such as user profile, item description, and reviews of users to items. Wang and Blei~\cite{wang2011collaborative} proposed the collaborative topic regression~(CTR)~that adopts the Latent Dirichlet Allocation~(LDA)~technique to improve the traditional collaborative filtering under a probabilistic framework. Afterward, several variants of CTR were presented, which also employed LDA to discover valuable aspects from textual reviews~\cite{ling2014ratings}. In addition, Bao {\it et al.}~\cite{bao2014topicmf} proposed a novel MF recommendation based on topic modeling, which employs the non-negative matrix factorization to derive topics from textual reviews. 

It should be mentioned that the aforementioned models mainly adopted the bag-of-words model and
ignored the contextual information of documents, e.g., the surrounding words and word orders. To address this issue, Kim {\it et al.}~\cite{kim2016convolutional} firstly
utilized deep learning~(DL) models to capture the contextual understanding of textual information, and proposed a novel document context-aware recommendation model called Convolutional Matrix Factorization~(ConvMF). Zhang {\it et al.}~\cite{zhang2017autosvd++} developed a new hybrid model that jointly models content information as representations of effectiveness and compactness, and leverages implicit user feedback to make accurate recommendations. Lu {\it et al.}~\cite{lu2018coevolutionary} presented an innovative recommendation model, which utilizes the attention-based recurrent neural networks to extract topical information from review documents. Although these models achieved outstanding performance, they ignored the natural selection bias problem caused by the potential distortion of available textual information, and thus cannot accurately reflect the target population. 

In this paper, we propose a novel and general self-adaptive module called the Self-adaptive Attention Module (SAM), which can self-adaptively learn attention by utilizing the representation of textual information to offset the selection bias. Our module can be seamlessly integrated into any models that contain learning components of textual information. In particular, the integrated model effectively takes advantage of the attention learned from the representation of textual information to enhance the prediction accuracy by altering the objective function of MF. We evaluate the effectiveness of SAM on three real-world datasets. Experimental results demonstrate that the integrated models can significantly outperform the state-of-the-art models. Moreover, extensive experiments verify that our proposal achieves performance improvement over a series of sparse scenarios.

The main contributions of this paper are summarized as follows.
\begin{itemize}
\item To the best of our knowledge, we are the first to propose a general module to alleviate the selection bias derived from utilizing available textual information in RS.
\item The proposed module can be seamlessly integrated into any recommendation models, including learning components of the textual information.
\item Experiments demonstrate the superiority of the integrated models with SAM and effectiveness over a series of sparse scenarios.
\end{itemize}
\section{Preliminary}
\subsection{Matrix Factorization}
Matrix factorization generate the latent features matrices for users and items, and obtain the predicted rating using inner product of user matrix and item matrix. We denote observed rating matrix as $\mathbf{R}_{\mathbf{\Omega}} \in \mathbb{R}^{m\times n}$ where $m$ and $n$ indicate the number of users and items, and $\mathbf{\Omega}$ refers to the indices set of observation. Suppose the true hidden rating matrix $\mathbf{R}_{\mathbf{O}}$ can be decompose into two latent feature matrices $\mathbf{U}\in \mathbb{R}^{m\times d}$ and $\mathbf{V}\in \mathbb{R}^{n\times d}$ of rank $d$. The rating $r_{ij}$ for user $i$ and item $j$ can be predicted by $\hat{r}_{i j}=\mathbf{u}_{i}^{\top} \mathbf{v}_{j}=\sum_{k=1}^{d} u_{ik} v_{kj}$.

To obtain the latent features $\mathbf{U}$ and $\mathbf{V}$, a general way is to minimize the loss function $\mathcal{L}$ which represents the sum of square loss between the observed ratings and corresponding predicted ratings. To avoid overfitting issue, $L_2$ regularization term tend to be considered into loss function as follows:
\begin{equation}
\begin{split}
    \mathcal{L}= \sum_{i,j}{I}_{ij}\left(r_{i j}-\mathbf{u}_{i}^{\top} \mathbf{v}_{j}\right)^{2}+\lambda_{u} \sum_{i}\left\|\mathbf{u}_{i}\right\|^{2}+\lambda_{v} \sum_{j}\left\|\mathbf{v}_{j}\right\|^{2},
\end{split}
\label{mf-loss-func}
\end{equation}
where $\mathbf{I}$ denotes indicator matrix satisfying $I_{ij}$ equals $1$ if $(i,j) \in \boldsymbol{\Omega}$ and equals $0$ otherwise, and $\left\|\cdot\right\|$ denotes the $L_2$ norm.

\subsection{Text Representation Learning}
In this part, we briefly review the DL models for text representation. Feed-forward networks are among the simplest DL models to achieve text representation. For example, Iyyer {\it et al.}~\cite{iyyer2015deep} introduced the Deep Averaging Network~(DAN), which feeds an unweighted average of word vectors through multiple hidden layers. Le and Mikolov~\cite{le2014distributed} proposed an unsupervised learning algorithm, Paragraph Vector, which learns vector representations for variable-length pieces of texts. These models essentially considered the text as a bad of words and learn a vector representation for each word while ignoring the word dependencies and text structures.

To address this issue, methods based on Convolutional Neural Network~(CNN) and Recurrent Neural Network~(RNN) have been developed to capture a deeper understanding of textual information. Kalchbrenner {\it et al.}~\cite{kalchbrenner2014convolutional} firstly proposed a CNN-based model for text representation. Afterward, Kim~\cite{kim-2014-convolutional} proposed a simpler CNN-based model which only has one layer of convolution and performs remarkably well. CNN-based models focus on patterns of words across space, whereas RNN-based ones can better capture time-level features of words. Tai {\it et al.}~\cite{tai2015improved} introduced a generalization of the standard Long Short-Term Memory~(LSTM) architecture to tree-structured network topologies, which achieves high performance for representing sentence meaning over a sequential LSTM. 

Considering huge success attention mechanism achieved in computer vision, it has been attracted increasingly researches in the text representation. Recently, there have been many works on incorporating attention module into various DL models such as~\cite{yang2016hierarchical}. In contrast to our proposal, these models focused on alleviating the bias from the perspective of model structure.

\subsection{Selection Bias}
It has been well known that if the observations do not represent the distribution of the underlying data, selection bias will appear~\cite{ovaisi2020correcting}. The reason may be that the research is adulterated with artificial selection criteria. There are many studies trying to correct selection bias. Heckman~\cite{heckman1979sample} discussed the bias that results from the usage of non-randomly selected samples. Smith and Elkan~\cite{smith2004bayesian} used Bayesian networks to formalize different types of selection bias. 
\section{Methodology}
 
 \begin{figure*}[!h]
	\centering
	 \includegraphics[width=0.8\linewidth]{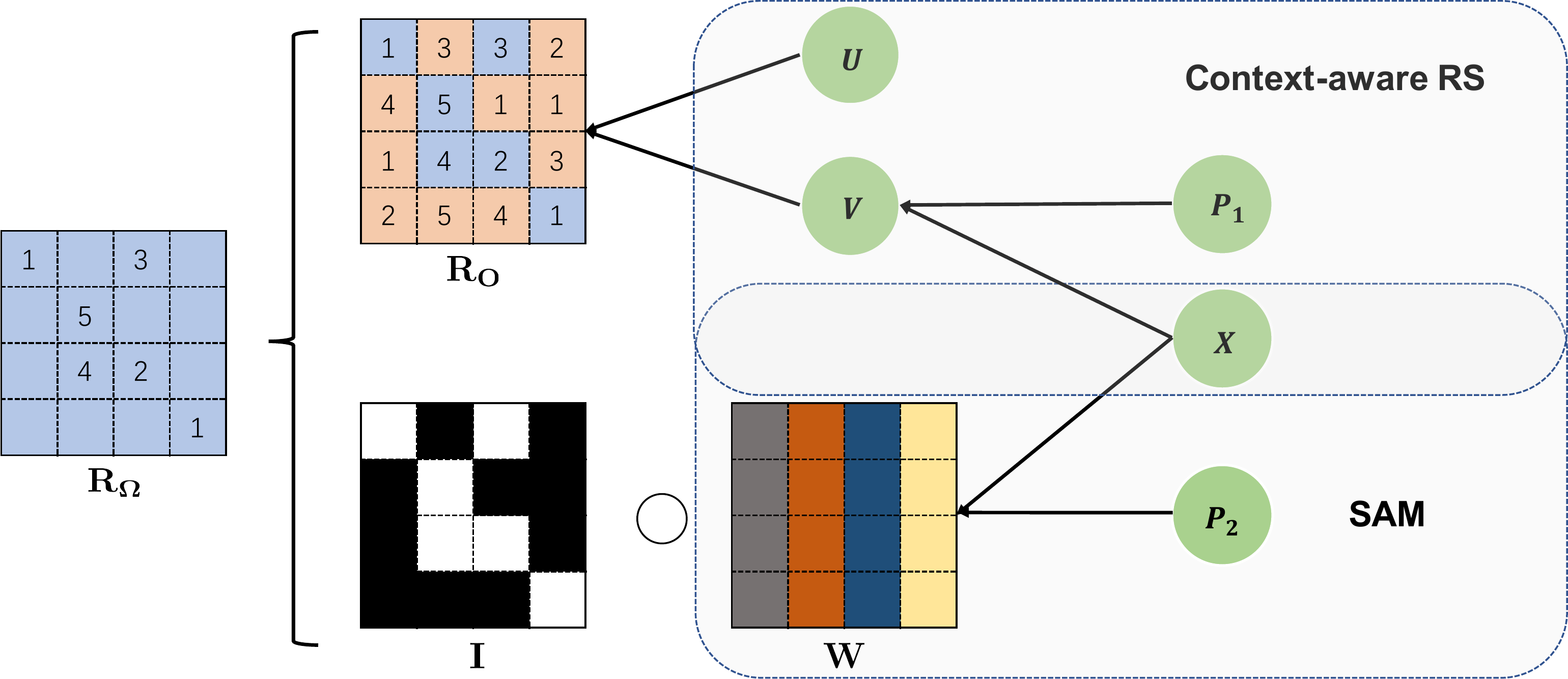}
    \caption{The architecture of integration of context-aware RS and SAM module. The observed rating matrix $\mathbf{R}_{\mathbf{\Omega}}$ consists of the completed matrix $\mathbf{R}_\mathbf{O}$ from recommendation models and Hardmard product~($\circ$) of the indicator matrix $\mathbf{I}$ and the bias-corrected weights matrix $\mathbf{W}$. The positions with numbers in $\mathbf{R}_{\mathbf{\Omega}}$ represent the observed ratings. In $\mathbf{R}_\mathbf{O}$, numbers in orange positions represent the true hidden ratings. The black and white areas in $\mathbf{I}$ represent the masked and observed positions, respectively. In $\mathbf{W}$, each column has a unique color indicating specific attention from SAM.}
	\label{arch}
\end{figure*}
 
 \subsection{Overview}

 The sketch of the integrated models is shown in Figure~\ref{arch}. Specifically, we decompose the observed matrix $\mathbf{R}_{\mathbf{\Omega}}$ into the completed matrix $\mathbf{R}_\mathbf{O}$ and the bias-corrected weights matrix $\mathbf{W}$. We denote textual information by $\mathbf{X}$, the internal parameters in the DL models of context-aware RS and SAM by $\mathbf{P_1}$ and  $\mathbf{P_2}$, respectively. The training of SAM is independent of $\mathbf{U}$ and $\mathbf{V}$ whereas $\mathbf{W}$ is indispensable to obtain $\mathbf{U}$ and $\mathbf{V}$. The natural idea is to model the probabilities of the selection process and adopt a bias-corrected risk function:
\begin{equation}
\mathcal{L}_{\mathbf{R}}=\sum_{i=1}^{m}\sum_{j=1}^n I_{ij}w_{ij}\left(r_{i j}-\mathbf{u}_{i}^{\top} \mathbf{v}_{j}\right)^{2},
\end{equation}
where $w_{ij}$ is the corresponding bias-corrected weights. With the consideration of auxiliary textual information, a naive modeling of the bias-corrected weights would be $w_{ij}=P^{-1}(I_{ij}=1|x_{j})$, where $x_{j}$ denotes the context information of item $j$. This leads to an unbiased risk function due to the observation $E(\mathcal{L}_{\mathbf{R}}|\mathbf{U},\mathbf{V})=\sum_{i=1}^{m}\sum_{j=1}^n (r_{i j}-\mathbf{u}_{i}^{\top} \mathbf{v}_{j})^{2}$. As $\{P(I_{ij}=1|x_{j})\}$ are usually unknown in practice, most works provided several parametric ways to model it and plugged the estimators of propensity scores into different rating estimation procedures respectively. However, these works suffer from the prior knowledge of specific selection model and usually requires additional parameter tunning process to avoid extreme propensities phenomena \cite{Rubin01,Kang-Schafer07}. We observed that the textual information influences the target ratings, more importantly, can interact with the rating process in most context-aware RS. To include the textual information in the selection model without the requirement of specific modeling, we adopt a DL-based method to obtain the bias-corrected weightings motivated by~\cite{kim2016convolutional}.

\subsection{DL-based Bias-corrected Weightings}
More specifically, We aim to choose the desired DL-based bias-corrected weights $\{w_{ij}\}$ for a range of $\mathbf{U} \in \mathbb{R}^{m\times d}$ and $\mathbf{V} \in \mathbb{R}^{n\times d}$ such that 
\begin{equation}
\sum_{i=1}^{m}\sum_{j=1}^n \left(I_{ij}w_{ij}-1\right)\left(r_{i j}-\mathbf{u}_{i}^{\top} \mathbf{v}_{j}\right)^{2}\approx0.
\end{equation}
To allow our weight can be adaptively embedded into different recommendation systems, we need to remove the specific rating information and one direct way is to control $\sum_{i=1}^{m}\sum_{j=1}^n (I_{ij}w_{ij}-1)^2$ under the constraint that all $w_{ij}\ge 1$. To adopt a DL-based bias-corrected weights, we propose the following optimization:
\begin{equation}
\min\left|\left|\mathbf{I}\circ \mathbf{W}(\mathbf{X})-\mathbf{J}\right|\right|_{\mathbf{S}},\quad\text{s.t.}\quad w_{ij}(x_j)\ge 1
\end{equation}
where $\mathbf{I}=(I_{ij})$ and $\mathbf{J}$ represents indicator matrix and matrix with all entries $1$ respectively, and $\circ$ denotes the elementwise Hardmard product of two matrices. We adopt the objective function as $||\cdot||_{\mathbf{S}}$ to be the largest singular value of a matrix to evaluate all the bias-corrected weights simultaneously. Note that the bias-corrected rates are restricted to be greater than or equal to $1$ due to the observation that $P^{-1}(I_{ij}=1|x_{j})$ should be greater than or equal to $1$. As for each weight $w_{ij}(x_j)$, we adopt a CNN architecture to learn $w_{ij}(x_j)=\text{cnn}(\mathbf{P_2}, x_j)$. By taking the regularization terms into consideration, we can obtain the complete loss function by extending $\mathcal{L}_{\mathbf{R}}$:
\begin{equation}
\begin{split}
    \mathcal{L} = \mathcal{L}_{\mathbf{R}} +\lambda_{u} \sum_{i=1}^{m}\left\|\mathbf{u}_{i}\right\|^{2} 
    +\lambda_{v} \sum_{j=1}^{n}\left\|\mathbf{v}_{j} - \text{dl}(\mathbf{P_1},  X)\right\|^{2},
\end{split}
\label{sam-loss-func}
\end{equation}
where $\text{dl}(\mathbf{P_1},  X)$ represents latent vectors of items via one specific DL model, detailed implementations of which can be obtained by referring to the following section.

\section{Empirical Study}
\subsection{Experimental Setting}
\subsubsection{Datasets}
We evaluate our proposal on two movie rating datasets from MovieLens~\cite{harper2015movielens} and the news reading-time dataset created by ourselves. For MovieLens datasets, users rate items on a scale of $1$ to $5$, which is similar to other homogeneous datasets. To verify the performance of our proposal over various recommendation scenarios, we build a new dataset that comes from Sogou News and consists of the reading-time of users on the news. To avoid problems caused by the scale of reading-time, we utilize the bi-scaling technique~\cite{bao2014topicmf} to standardize the interaction matrix before training. The statistical description of datasets is shown in Table~\ref{stat}. 

\begin{table}[h!]
	\caption{Statistical description on three real-world datasets}
	\begin{center}
		\begin{tabular}{ccccc}
			\hline
			\text{Model}	&	\# \text{User}		&	\#\text{Item}	&	\#\text{Interaction}	&	Density\\
			\hline
			ML-100K		&	$943$		&		$1483$	&		$88673$	&	$6.341\%$\\
			ML-1M		&	$6040$	&	$3544$	&		$993482$		&	$4.641\%$\\
			Sougou News		&	$1728$	&	$2573$	&	$52912$		&	$1.190\%$\\
			\hline
		\end{tabular}
	\end{center}
	\label{stat}
\end{table}

Note that MovieLens does not contain the textual information, we extract the movie descriptions from IMDB and perform preprocessing on it as~\cite{kim2016convolutional} did. For Sougou News, we set the length of news to $500$ words by truncating the longer news and padding the shorter news. In addition, we select the top $36000$ words as a vocabulary and remove all non-vocabulary words from the news.

\begin{table}[h]
	\caption{Overall test RMSE. Models with '+' in their names represent combinations with SAM.}
	\begin{center}
		\begin{tabular}{cccc}
			\hline&\multicolumn{3}{c}{\textbf{Dataset} }  \\ 
				\cline{2-4}
				\textbf{Model} & \text{ML-100K} & \text{ML-1M} & \text{Sougou News}\\
			\hline
				\text{MF}            &     0.959	&	0.902	&	\textcolor{blue}{0.940}\\
				\text{ConvMF}	&	\textcolor{blue}{0.910}		&	0.853	&	0.954\\
				\text{FTMF}&	0.928	&	\textcolor{blue}{0.845}	&	0.961\\
				\text{RCNNMF}&	0.931		&	0.846	&	0.966\\
			\hline
				\text{MF+}            &     0.958	&	0.880	&	\textcolor{red}{0.938}\\
				\text{ConvMF+}	&	\textcolor{red}{0.907}		&	0.844	&	0.939\\
				\text{FTMF+}&	0.922	&	\textcolor{red}{0.840}	&	0.948\\
				\text{RCNNMF+}&	0.915		&	0.843	&	0.961\\
			\hline
			\text{Improve} & $0.33\%$ & $0.59\%$ & $0.17\%$\\
			\hline
		\end{tabular}
	\end{center}
	\label{rmse}
\end{table}

\subsubsection{Competitors}

We employ the classical MF and ConvMF, a representative model combining MF with DL models for textual information as baselines. Considering that ConvMF provides a framework for integration of MF and DL models, we design two new models RCNNMF that replaces ``Conv'' part with Recurrent Convolutional Neural Networks~(RCNN)~\cite{lai2015recurrent}, and FTMF that substitutes FastText~\cite{joulin2016bag} for ``Conv'' part for baselines. We evaluate the performance improvement produced by the integration of these models and the SAM.

\subsubsection{Evaluation Metric}
For each dataset, we randomly split observations into $80\%$ and $20\%$ as train/test sets. To make MF working on all users and items, we ensure at least one interaction existing on each user and item. We evaluate the performance of prediction using the Root Mean Square Error (RMSE) metric, which is computed by $\mathrm{RMSE}=\sqrt{\sum_{\mathbf{\widetilde{\Omega}}}\left(r_{i j}-\hat{r}_{i j}\right)^{2} / |\mathbf{\widetilde{\Omega}}|}$ where $\mathbf{\widetilde{\Omega}}$ indicates the indexes of test set. To avoid the overfitting, we set the maximum iteration to $200$ with early-stopping.

\subsection{Performance Comparison}

Table~\ref{rmse} shows the performance of baselines and models with SAM. Considering the range of Sougou News, we scale the results by multiplying $10^{-2}$ for performance display. We observe that models with SAM significantly outperform the corresponding baselines on all datasets. Specifically, for two MovieLens datasets, the best performances are achieved by integrating ML and DL models. The improvements demonstrate the effectiveness of SAM when textual information facilitates the rating prediction. For Sougou News, ML with SAM also obtains considerable improvement over the vanilla MF, which further proves the potency of SAM.

\begin{table}[h!]
	\caption{Test RMSE over various sparseness of training data on MoviesLens $100$K dataset. Models with '+' in their names represent combinations with SAM.}
	\begin{center}
		\begin{tabular}{cccc}
			\hline&\multicolumn{3}{c}{\textbf{Training ratio of dataset~(density)}}  \\ 
			\cline{2-4}
			\textbf{Model} & $20\%$($1.27\%$) & $40\%$($2.54\%$) & $60\%$($3.80\%$)	\\
			\hline
			\text{MF}            &     $1.248$	&	$1.070$	&	$1.001$ \\	
			\text{ConvMF}	&	\textcolor{blue}{$1.044$}		&	\textcolor{blue}{$0.962$}	&	\textcolor{blue}{$0.928$}	\\
			\text{FTMF}&	$1.115$	&	$0.989$	&	$0.972$	\\
			\text{RCNNMF}&	$1.163$	&	$1.011$	&	$0.971$		\\
			\hline
			 \text{MF+}          &     $1.223$	&	$1.063$	&	$1.000$	\\
			\text{ConvMF+} 	&	\textcolor{red}{$1.020$} &	$0.957$	&	\textcolor{red}{$0.927$}	\\
			\text{FTMF+}  &  $1.058$	&	$0.987$	&	$0.955$	\\
			\text{RCNNMF+} & $1.021$ &	\textcolor{red}{$0.951$}	& $0.930$\\
			\hline
			\text{Improve} & $2.30\%$ & $1.14\%$ & $0.11\%$\\
			\hline
		\end{tabular}
	\end{center}
	\label{ratio_rmse}
\end{table}

Table~\ref{ratio_rmse} reveals that the comparison results of all methods under a series of missing scenarios. Overall, we observe that improvements caused by SAM get more significant when the train set becomes more sparse. It implies that SAM has excellent potential in extremely sparse scenarios.

\section{Conclusions}

In this paper, we propose a general attention-based module SAM to alleviate selection bias derived from the utilization of textual information in recommendation systems, which can be seamlessly integrated into models containing learning components of the textual information. Empirical studies on three real-world datasets demonstrate the effectiveness of SAM, and extensive experiments imply the great potential of SAM under extremely sparse scenarios.

    \bibliographystyle{IEEEbib}
  \bibliography{bibfile}
\end{document}